\documentstyle [11pt,epsf] {article}

\begin{document}
\begin{titlepage}
\begin{center}
\vspace{2cm}
\LARGE 
The Clustering of Galaxies around Quasars 
\\                                                     
\vspace{1cm} 
\large
Guinevere Kauffmann$^1$ \& Martin G.  Haehnelt$^{2}$ \\
\vspace{0.5cm}
\small
{\em $^1$  Max-Planck Institut f\"{u}r Astrophysik, D-85740 Garching, Germany} \\
{\em $^2$  Astrophysics Group, Imperial College of Science, Technology and Medicine,
 Prince Consort Road, London, SW7 2BW, UK}\\
\vspace{0.8cm}
\end{center}
\normalsize
\begin {abstract}

We study the cross-correlation between quasars and galaxies  
by embedding models for the formation and evolution of the two populations in
cosmological N-body simulations. We adopt the quasar evolution model of 
Kauffmann \& Haehnelt (2000), in which supermassive black holes are formed 
and fuelled during major mergers.  We define the ``bias'' 
parameter $b_{\rm QG}$ as the ratio of the cross-correlation function $\xi_{\rm QG}$
to the galaxy auto-correlation function $\xi_{\rm GG}$.
On scales larger than 1 $h^{-1}$ Mpc, the values     
of $b_{\rm QG}$ predicted by our models at low redshift depend 
very little on galaxy selection.             
They measure the characteristic mass 
of the dark matter halos that host quasars and can be used to estimate the typical
quasar  lifetime. In current redshift surveys, such measurements will 
constrain the lifetimes of low redshift quasars more accurately
than measurements of the quasar autocorrelation function, because
galaxies have much higher space densities than quasars.
On scales smaller than
1 $h^{-1}$ Mpc, the main contribution to  
$\xi_{\rm QG}$ comes 
from quasar/galaxy pairs in the same dark matter halo. 
The amplitude of $\xi_{\rm QG}$  depends both on
the location of the host galaxy and on the density profile of other galaxies 
within the halo. As a result, measurements on these scales yield information about
the processes responsible for fuelling supermassive black holes.
At high redshifts our models predict that  
quasars of fixed luminosity are located in less massive halos than at low redshift.
They are therefore less biased relative to 
galaxies of given luminosity or stellar mass.
We have used the simulations to  calculate the evolution
of the quasar auto-correlation function. We find that                     
models with quasar lifetimes in the range 
$10^{6}-10^{7}$ years provide a good match to the results of
the 2dF QSO survey.

\end {abstract}
\vspace{1.3cm}
Key words:galaxies:formation -- galaxies:nuclei -- quasars:general -- black hole physics
\end {titlepage}

\section {Introduction}          
It has been known for more than three decades that quasars are associated with enhancements in
the distribution of galaxies (Bahcall, Schmidt \& Gunn 1969). Calculations of the QSO/galaxy
correlation functions on small scales ($< 1$ Mpc) have shown that at redshifts $z < 0.4$,
quasars typically reside in small to moderate groups of galaxies and not in rich
clusters (e.g. Hartwick \& Schade 1990; Bahcall \& Chokshi 1991; Fisher et al. 1996).

It is considerably more uncertain if the environments of quasars depend on their optical or radio
luminosities, or on redshift. Early studies claimed that radio-loud quasars were located
in richer environments than radio-quiet quasars (e.g. Yee \& Green 1984, 1987; Ellingson,
Green \& Yee 1991), but these findings have been contested in a number of recent papers
(e.g. Fisher, Bahcall \& Kirhakos 1996; McLure \& Dunlop 2001; Wold et al 2001).
Most studies have not found any relation between the optical luminosity of a quasar 
and its clustering (Brown, Boyle \& Webster 2001; Finn, Impey \& Hooper 2001), but
there have been a number of papers claiming the existence of a correlation 
between radio luminosity and
environment (e.g. Wold et al 2001; Finn, Impey \& Hooper 2001). The studies of Yee \& Green (1987)
and Ellingson, Yee \& Green (1991) found that high-redshift, radio-loud quasars were located in
richer environments than low-redshift, radio-loud quasars, but more recent analyses 
(e.g. Wold et al 2001)
fail to find any trend with redshift.

Part of the reason for the discrepancies between different studies may be that the quasar 
samples studied so far are small (typically 20-70 objects) and are selected in widely varying    
different ways. This situation will soon improve with the availability of a new generation
of very large quasar surveys. The 2dF QSO redshift survey (Croom et al 2001a) will
measure redshifts for $\sim$25000 optically selected quasars with $b_j < 20.85$ and $z<3$. 
The Sloan Digital Sky Survey (York et al 2000) aims to study $\sim 100 000$ quasars brighter
than $g' \sim 19.7$. The evolution of the redshift-space correlation function of quasars
in the 2dF survey has been measured by Croom et al (2001b).
Their study shows that the clustering of quasars in the survey is very comparable to
that of $L_*$ galaxies at low-redshift,  and that the correlation length $R_0$  remains
approximately constant out to $z \sim 2$.

In order to interpret observational results on the clustering properties of quasars, 
a physical model for the evolution of the quasar
population in a cosmological context is required (Haehnelt \& Rees 1993). 
Recently, we (Kauffmann \& Haehnelt 2000, hereafter KH; Haehnelt \& Kauffmann 2000)
introduced a ``unified'' model for the evolution of galaxies and quasars in a
cold dark matter (CDM) dominated Universe. 
We assumed that supermassive black holes were formed and 
fuelled during major mergers. If two galaxies of comparable mass merged, their central
black holes coalesced and a few percent of the gas in the merger remnant was accreted by
the new black hole over a timescale of a few times $10^7$ years. KH showed that
their model could reproduce quantitatively the observed relation between bulge
velocity dispersion and black hole mass in nearby galaxies, the strong evolution of the quasar
population with redshift and the observed relation between the luminosities of quasars
and their host galaxies.

In this paper, we extend the model of KH to study the galaxy environments and the clustering
properties of quasars as a function of optical luminosity and of redshift.
To do this, we combine the KH quasar evolution model with
 a set of cosmological N-body simulations in which galaxy formation
is included using simple recipes from standard semi-analytic
models (Kauffmann et al 1999a). 
This provides a fully spatially resolved simulation of the clustering of both
galaxies and quasars at a series of redshifts. In section 2, 
we briefly review the techniques used for
following quasar evolution in N-body simulations and we discuss the relation
between quasars and dark matter halos in the simulation. In section 3, we calculate
the clustering of galaxies around quasars at low redshift. In section 4, we predict
how quasar/galaxy correlations should evolve to high redshift. In section 5, we 
calculate the evolution of the quasar auto-correlation function and in
section 6, we summarize our results.

\section {Including Quasars and Galaxies in Cosmological N-body Simulations}

The techniques used to include galaxy formation in cosmological N-body simulations
are described in detail in Kauffmann et al (1999a). This paper also describes the global
properties of galaxies at $z=0$ in the simulation, including the luminosity function
and the two-point correlation function. The evolution of galaxy clustering to high redshift
is discussed in Kauffmann et al (1999b).

Briefly, a friends-of-friends group-finding algorithm is used to identify virialized
dark matter halos in the simulation at a series of closely spaced redshifts from $z\sim 20$
to the present. A halo merging tree is constructed, which links halos at $z=0$ to their
progenitors at each earlier output time. The galaxy formed from gas cooling in a dark matter           
halo is assigned the index of the most bound particle in that halo and is referred
to as the ``central galaxy'' of that halo. The galaxy remains identified with the same particle
when its halo is accreted by a larger group or cluster and it becomes a ``satellite''.
A satellite can merge with the central galaxy on a dynamical friction timescale. 
If the ratio of the masses
of the merging satellite and the central galaxy is greater than 0.3, we call this
a ``major merger'' and an elliptical merger remnant is formed. Following KH,
a fraction $f_{BH}$ of the available gas is accreted by the black hole and the remainder
will be converted into stars over a short ($10^8$ year) timescale in a ``burst''. 
Note that in our scheme, the quasar is always located at the position of the
central galaxy of the halo. In section 3, we will explore what happens if we
relax this assumption.

In this paper, we study a $\Lambda$CDM model with $\Omega=0.3$, $\Lambda=0.7$, $\sigma_8=0.9$
and $H_0= 70$ km s$^{-1}$ Mpc$^{-1}$.  The simulation (one of the two analyzed in 
Kauffmann et al 1999a)  contains 17 million particles of 
mass $2 \times 10^{10} M_{\odot}$ in a periodic box of  $L= 141 h^{-1}$ Mpc.
We will restrict our analysis to quasars that form in halos more massive than
$2 \times 10^{12} M_{\odot}$ (i.e. containing at least  100 particles), 
for which the simulation is able to provide a rough merging history.

We have checked that when we parametrize the fraction of gas
accreted by a black hole during a merger in the same way as in our previous
work using analytic merger trees, 
we obtain a relation between black
hole mass and bulge velocity dispersion with the same zero point. 
The  relation derived from the simulations exhibits considerably more scatter,
because the merging histories of halos 
are not followed as accurately in the simulation.
Merger trees derived using analytic methods 
(e.g. Kauffmann \& White 1993) are not limited
by mass resolution, but have disadvantage that these are not able to specify the 
positions of halos and how they evolve with time.
If we restrict our analysis to halos containing at least a thousand particles,
we obtain a tighter relation  
that is in much better agreement with the analytic models, 
but we are then only able to
follow the evolution of very bright quasars and there 
are not many such objects in the simulation, particularly at high redshifts.
We have adopted the 100 particle limit as a compromise,
and we caution that higher resolution simulations are needed to address the scatter
around the mean relations that we present in this paper. We also focus on
the clustering properties of quasars at $z< 2$, where there are
still plenty of halos more massive than 100 particles in the simulations.
Simulations of  larger
volumes at the same resolution are required in order to extend our analysis to
redshifts beyond 2. 

The KH model assumes that quasar activity is triggered by major merging events.
In practice, a variety of different physical mechanisms, including the accretion of small
gas-rich satellites, tidal interactions between galaxies, and the accretion of gas in
cooling flows, may also
result in quasar activity. Recent observations (McLure et al 1999)
suggest that at low redshifts, very luminous quasars with R-band magnitudes brighter than $-24$
are found  almost exclusively in elliptical host galaxies, but that fainter quasars 
are found in spiral as well as elliptical  hosts. This suggests that the KH
model is appropriate for  luminous quasars, but not for Seyferts
and other low-level AGN. At high redshifts, much less is known about the properties 
of quasars and their host galaxies, but it is likely that 
there are a variety of fuelling mechanisms.

In order to predict the optical luminosities of the quasar, KH assumed a relation            
between the mass of gas accreted by the black hole $M_{acc}$  and the absolute B-band magnitude
of the quasar at the peak of its light curve
\begin{equation} M_B (\rm{peak}) = -2.5 \log (\epsilon_B M_{acc}/ t_q) -27.45, \end {equation}
where $\epsilon_B$ is a radiative efficiency constant and $t_q$ is the quasar lifetime.
They also assumed that the luminosity of the quasar declined exponentially after
the merging event as
\begin {equation} L_B(t) = L_B (\rm{peak}) exp (-t/t_q). \end {equation}
KH considered a range of values of $t_q$ and concluded that a value $\sim 10^7$ years
provided the best fit to the shape of the observed quasar luminosity function.

In Fig. 1, we plot the mass distributions of dark matter halos that host quasars
with absolute B-band magnitudes of -23.5 (solid lines) and -25 (dashed lines)
for three different choices of the quasar lifetime $t_q$. Results are shown at $z=0.4$
and at $z=1.7$. Note that here and in the rest of this paper, the value
of the quasar lifetime $t_q$ is always quoted at $z=0$.  
KH assumed that $t_q$ scaled with the host galaxy dynamical time,  i.e.
$\propto [0.7 +0.3(1+z)^{3}]^{-1/2}$ in a  $\Lambda$CDM cosmology.
This means that $t_q$ is a factor of two smaller at $z=1$ and a factor of
3 smaller at $z=2$.

The mass distribution of the host halos  
is very sensitive to $t_q$. Large values of $t_q$ mean that a given mass of accreted gas
will have a lower peak luminosity (equation 1). As a result, 
bright quasars  occupy massive halos,
which contain galaxies with large gas masses.  
In the KH model, where a fixed fraction of the available gas is accreted by
the black hole during each merging event and where
the peak quasar luminosity scales  with                  
the mass of the halo, long quasar lifetimes result in a very steep
quasar luminosity function,  
because quasars sample the exponentially declining high-mass end  
of the halo mass function.  Small values of $t_q$ place quasars in less massive halos,
where the halo mass function is comparatively flat. As a result, the luminosity function 
has a much shallower
dependence on magnitude. However, as discussed by Haehnelt, Natarajan \& Rees (1998), a more
complicated fuelling prescription would lead to different inferred lifetimes. 

Fig. 2 compares the quasar luminosity functions obtained
for three different values of $t_q$ with new data from the 2dF Quasar survey.
We plot the luminosity function derived from the 10k catalogue
(Croom et al 2001a) for a cosmology with $\Omega=0.3$ and   
$\Lambda=0.7$. 
We find that  values of $t_q $ in the range  $10^6 - 10^7$ years appear to
provide the best fit to the data. Note however, that the differences between
the models are largest close to the magnitude limit of the survey, where the data is
most subject to photometric incompleteness (Boyle et al 2000). It is thus
difficult to assess whether the apparent ``turnover'' in the luminosity function
at $M_B > -23$ in the z=0.4 luminosity function and at
$M_B > -24$ in the z=0.9 function  is real or is caused by selection effects.
In the next section, we show that the clustering amplitude  of galaxies around bright
quasars also constrains the value of $t_q$ in our models. Unlike the luminosity
function, the clustering amplitude measures the mass distribution of the dark
matter halos that host quasars of given luminosity, 
so the derived lifetimes do not depend on the
adopted fuelling prescription.

\begin{figure}
\centerline{
\epsfxsize=11cm \epsfbox{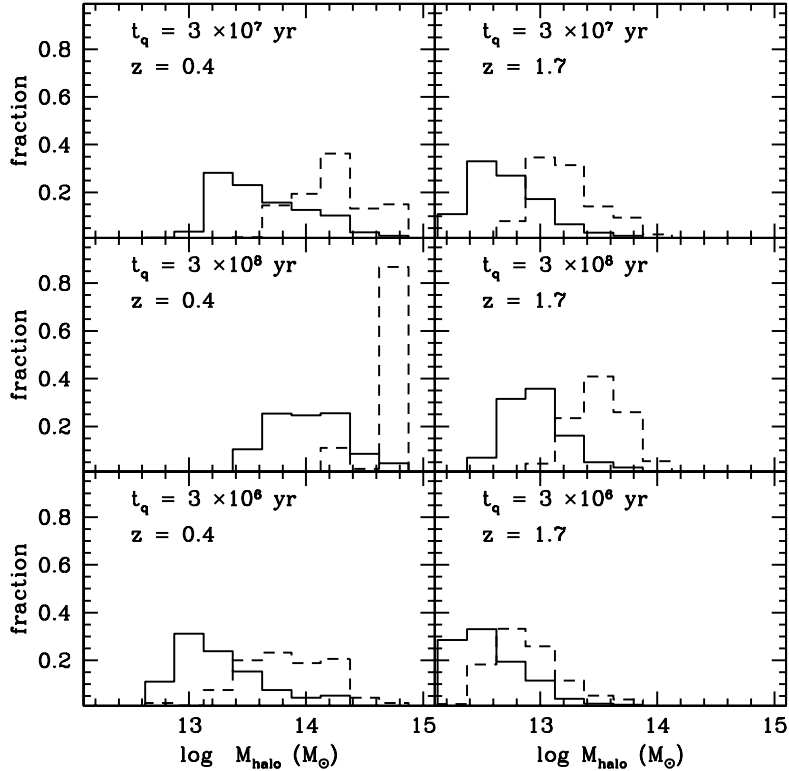}
}
\caption{\label{fig1}
\small
The mass distribution of dark matter halos that contain quasars with absolute B-band magnitude     
$-23.5$ (solid lines) and $-25$ (dashed lines) for three different values of $t_q$ at z=0.4 and 
z=1.7.}
\end {figure}
\normalsize

\begin{figure}
\centerline{
\epsfxsize=10cm \epsfbox{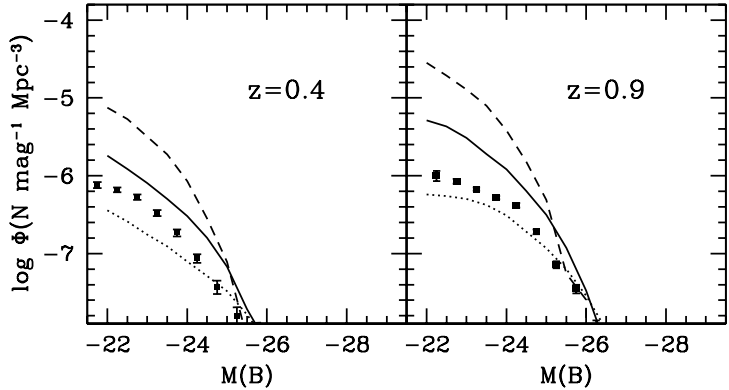}
}
\caption{\label{fig2}
\small
The quasar luminosity function at $z=0.4$ and $z=0.9$. The different lines are for different
values of the quasar lifetime, $t_q$: $3 \times 10^7$ yr (solid), $3 \times 10^{6}$ yr (dotted)
and $3 \times 10^{8}$ yr (dashed).
The solid symbols show results from the 2dF 10k catalogue for 
a cosmology with $\Omega=0.3$ and $\Lambda=0.7$ (Croom et al 2001a).} 
\end {figure}
\normalsize

\section {Quasar/Galaxy Correlations at Low Redshift}

Analyses of N-body simulations demonstrate that the correlation function of dark matter halos
is proportional to that of the mass over a wide range in scale. For massive halos, 
the constant of proportionality
is well predicted by a simple analytic model based on an extension of the Press-Schechter
theory (Mo \& White 1996). More recently, Sheth, Mo \& Tormen (2001) have shown that
an improved model, based on the assumption that halos form in an ellipsoidal
rather than a spherical collapse, can explain the clustering properties of
low mass halos. In general, more massive halos are clustered more strongly than less massive
halos. For CDM models, the dependence of clustering on halo mass is 
weak for halos with masses below $\sim 10^{13} M_{\odot}$, but stronger for             
more massive halos (Jing 1998). 
This is illustrated in the left-hand panel of Fig. 3, where we plot the
correlation function of halos in different mass ranges in our simulation at $z=0.4$.
At this redshift, halos less massive than a few times  $10^{13} M_{\odot}$ are clustered         
more weakly than the underlying dark matter, and halos more massive than this are 
clustered more strongly than the dark matter.

In the KH model, quasar activity occurs only in halos in which the central
galaxy has experienced a recent major merger. One might ask whether these
halos are clustered any differently from the average halo of the same mass. 
The right-hand panel of Fig. 3 compares the auto-correlation function
of all halos more massive than $2 \times 10^{12}$ $M_{\odot}$  with
the result obtained when halos containing quasars are                                   
cross-correlated with this sample. 
Results are shown at $z=1$, but the conclusion is the same at all redshifts.
The clustering of quasar host halos  is indistinguishable
from that of ordinary halos of the same mass, except on scales less than $\sim 1 h^{-1}$ Mpc,
where they are clustered somewhat more strongly.
Note that the correlation function falls off steeply on scales smaller than 
$1 h^{-1}$ Mpc, because this corresponds roughly to the virial radius of the typical
halo in our sample.

\begin{figure}
\centerline{
\epsfxsize=9cm \epsfbox{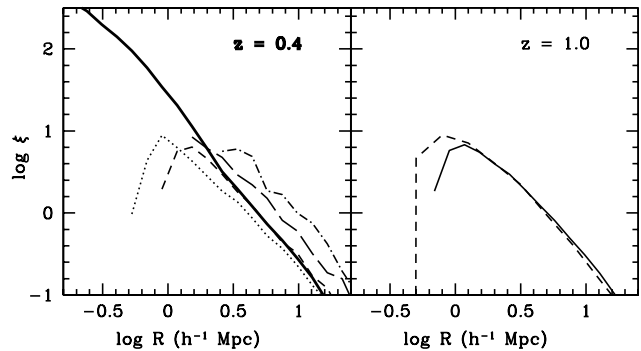}
}
\caption{\label{fig3}
\small
{\bf Left}: The thick solid line shows the correlation function of the dark matter at $z=0.4$.
The thin lines show the halo correlation functions for halos in the mass range
$10^{12}$ -- $3 \times 10^{12}$ $M_{\odot}$ (dotted), 
$3 \times 10^{12}$-- $10^{13}$ $M_{\odot}$ (short-dashed),  
$10^{13}$ -- $3 \times 10^{13}$ $M_{\odot}$ (long-dashed)), 
and $3 \times 10^{13}$-- $10^{14}$ $M_{\odot}$ (dashed-dotted).  
{\bf Right}: The solid line shows the halo auto-correlation function for halos more massive
than $2 \times 10^{12}$ $M_{\odot}$ in the simulation.
The dashed line shows the result obtained when  halos 
containing quasars are cross-correlated with the
full sample.}                  
\end {figure}
\normalsize

The left hand panels of Fig. 4 show quasar/galaxy cross-correlation 
functions $\xi_{\rm QG}$ for our ``fiducial'' model with 
$t_q = 3 \times 10^7$ yr. The thin lines on the diagram illustrate the results obtained for
quasars of different luminosities, and the thick solid line shows the galaxy auto-correlation
function for comparison. We select galaxies from the simulation in two different ways:
by stellar mass and by B-band absolute magnitude. We find that the selection procedure
strongly influences the amplitude of the correlation functions on small scales.
Galaxies selected by stellar mass are more strongly clustered on scales less than
$1 h^{-1}$ Mpc than galaxies selected by B-band absolute magnitude.
The right hand panels of Fig. 4 indicate the contribution to $\xi_{\rm QG}$
from quasar/galaxy pairs in the {\em same dark matter halo}. As can be seen, 
the contribution to $\xi_{\rm QG}$ from quasars and galaxies in the same halo dominates
on scales smaller than $ \sim 1 h^{-1}$ Mpc. The amplitude and slope of $\xi_{\rm QG}$ 
depend on the density profile of galaxies within individual dark matter halos.
We note that the galaxy  density profiles of clusters in our simulation               
have been compared with observational data and have been found to be in good
agreement (Diaferio et al 2001; Springel et al 2001).

\begin{figure}
\centerline{
\epsfxsize=11cm \epsfbox{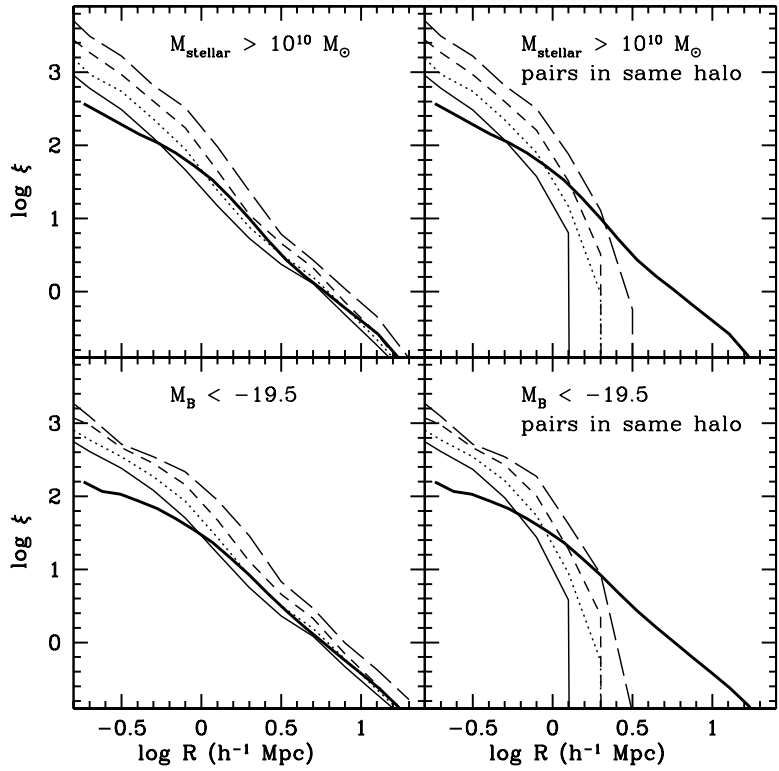}
}
\caption{\label{fig4}
\small
{\bf Left}: Thin lines show quasar/galaxy cross-correlation functions for quasars with B-band
absolute magnitude of $-22.5$ (solid), $-23.5$ (dotted), $-24.5$ (short-dashed) and $-25.5$ 
(long-dashed). The thick solid line shows the galaxy auto-correlation function.
{\bf Right}: Thin lines show quasar/galaxy cross-correlation functions for quasar/galaxy
pairs in the same dark matter halo. All results are for our fiducial model with
$t_q = 3 \times 10^7$ yr at a redshift $z=0.4$.}
\end {figure}
\normalsize

We have assumed that the quasar is always located at the position of 
the central galaxy in the halo.
Simulations by Mihos \& Hernquist (1996) show that during the merging of two disk galaxies
of near-equal mass, gas inflows may occur at more than one stage during the encounter.
After the initial collision, both galaxies form a strong central bar, and this
drives an inflow of gas  while the two galaxies are still widely separated.
When the two galaxies finally merge, the remaining gas is again driven towards the
centre of the remnant. The amount of gas channelled towards the centre during
each event depends on the geometry of the encounter and on whether the galaxies have
central bulges or not. These details of the fuelling process will affect the correlation
functions on scales where tidal interactions between galaxies
become important. Kochanek, Falco \& Munoz (1999) have pointed out that  
if fuelling occurs during galaxy encounters, the amplitude of the quasar autocorrelation 
function on small scales would  be enhanced over a simple power-law
extrapolation from large radii. 
We do not address these issues in our simulations, but Fig. 5
illustrates what happens if we relax the assumption that quasars are always central objects
and allow the quasar to be located at the position of a random galaxy in the halo.  
$\xi_{\rm QG}$ is now much shallower on scales smaller than $\sim 1 h^{-1}$ Mpc.

\begin{figure}
\centerline{
\epsfxsize=10cm \epsfbox{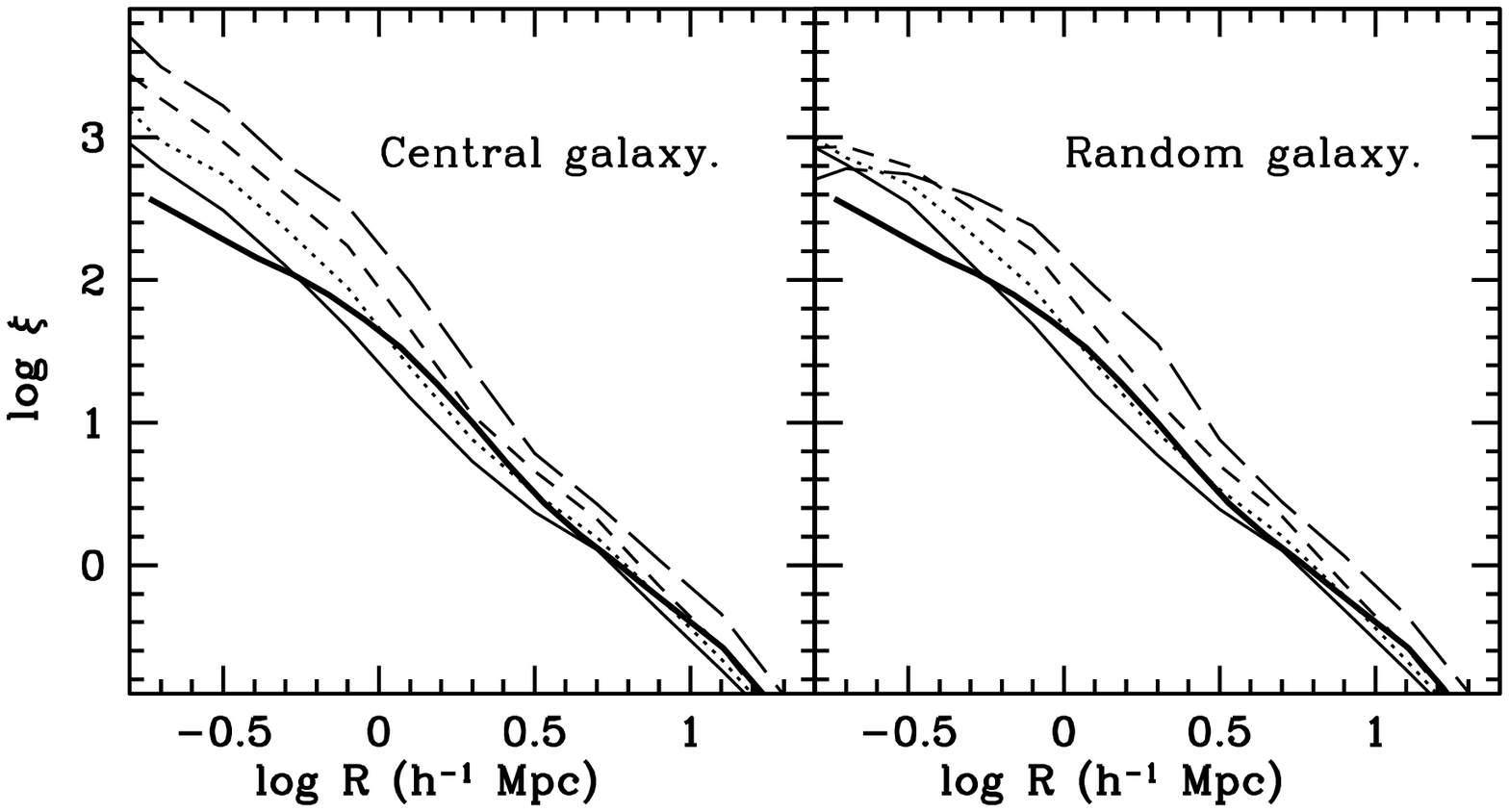}
}
\caption{\label{fig5}
\small
{\bf Left}: Quasar/galaxy cross-correlations for quasars located at the centres of dark matter halos.
Lines are as in Fig. 4.
{\bf Right}: Quasar/galaxy cross-correlations for quasars located at the positions of random
galaxies in the halo.
Results are for our fiducial model with
$t_q = 3 \times 10^7$ yr at a redshift $z=0.4$.}
\end {figure}
\normalsize

It is useful to define the quantity $b_{\rm QG}(R)$  as the ratio of the amplitude of the quasar/galaxy
cross-correlation function $\xi_{\rm QG}(R)$ to that of the galaxy auto-correlation function
$\xi_{\rm GG}(R)$:
\begin{equation} b_{\rm QG}(R) = \xi_{\rm QG}(R) / \xi_{\rm GG} (R). \end {equation}
Kauffmann, Nusser \& Steinmetz (1997) have shown that when 
smoothed on scales larger than $\sim 1 h^{-1}$ Mpc,
the relation between the galaxy and dark matter density fields in 
N-body simulations is well described
by a linear biasing relation for galaxies of all types and luminosities, so that
$\delta_{\rm G} = b_{\rm G} \delta_{\rm DM}$. Likewise, the quasar density field $\delta_{\rm Q}$ may be
written $\delta_{\rm Q} = b_{\rm Q} \delta_{\rm DM}$.  The quantity $b_{\rm QG}$ thus 
measures the ratio $b_{\rm Q}/b_{\rm G}$, i.e.  the relative bias of quasars and galaxies.

The behaviour of $b_{\rm QG}$ as a function of radius is shown in Fig. 6 
for galaxy samples selected in a variety of different ways.
We see that on scales larger than $\sim 1 h^{-1}$ Mpc , $b_{\rm QG}(R)$   
depends very little on the way in which galaxies are selected in the simulation. 
This is not surprising. As we have noted, in CDM cosmologies 
the clustering amplitude of present-day halos is weakly dependent
on mass for halos less massive than $\sim 10^{13} M_{\odot}$. This corresponds to 
the characteristic mass ($M_*$) of non-linear objects at the present day. As discusssed
by Jing (1998), halos smaller than this are almost   
unbiased tracers of the underlying dark matter distribution.   On the other hand,
the clustering amplitude is a strong function of mass for halos more massive than
than $M_*$. Because  low-redshift $L_*$ galaxies are located  primarily in halos
of $\sim 10^{12} M_{\odot}$ and quasars are found  in halos more massive 
than $10^{13} M_{\odot}$ for all values of $t_q$ (Fig. 1),
$b_{\rm QG}$ is a sensitive probe of mass distribution of the halos
hosting quasars, but not of the masses of the halos that contain galaxies.
In Fig. 6, brighter quasars appear more ``biased'' than fainter quasars 
because they are located in  more massive dark matter halos. 
We note that the bias parameter will be considerably more sensitive
to galaxy selection at high redshifts, where galaxies bright 
enough to be detected are located 
in halos more massive than $M_*$ at that epoch.

\begin{figure}
\centerline{
\epsfxsize=8.5cm \epsfbox{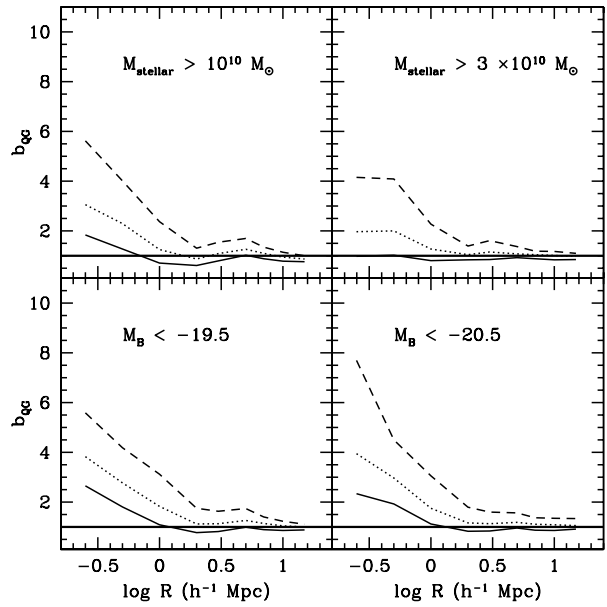}
}
\caption{\label{fig6}
\small
$b_{\rm QG}(R)$ for galaxy samples selected in different ways in the simulation.
Solid, dotted and dashed lines show results for quasars with $M_B =$ -22.5,
-23.5 and -24.5.
Results are for our fiducial model with
$t_q = 3 \times 10^7$ yr at a redshift $z=0.4$.}
\end {figure}
\normalsize

Fig. 7 demonstrates that $b_{\rm QG}(R)$ is strongly dependent 
on the quasar lifetime $t_q$. For long lifetimes, 
there are large predicted  differences in the clustering amplitudes
of bright and faint quasars, particularly on scales between 1 and 2 $h^{-1}$ Mpc.
For short lifetimes ($10^{6}$ yr), the differences between the clustering amplitudes of bright and faint
quasars are small. Our fiducial model is intermediate between
these two cases.
Low redshift quasars with $M_B = -24.5$ are predicted to be clustered $\sim 1.5$ times
more strongly than quasars with $M_B = -22.5$ on scales larger than 3 $h^{-1}$ Mpc.
On scales $\sim 1 h^{-1}$ Mpc, the difference is predicted to be closer to a factor of three, but
this will depend on the location of the quasar in the halo.

It has been suggested previously (La Franca et al 1998; Haehnelt, Natarajan \& Rees 1998; 
Haiman \& Hui 2001; Martini \& Weinberg 2001) that
the amplitude of the quasar auto-correlation function  constrains 
the quasar lifetime. Both $\xi_{\rm QQ}$ and $\xi_{\rm QG}$ probe the masses of quasar
host halos. In practice,  $\xi_{\rm QG}$  provides a stronger constraint
at low redshifts where the number density of quasars is very low. 
As seen in Fig. 7, the main difference between models with short
($t_q \sim 10^6$ yr) and long ($t_q \sim 10^8$ yr) lifetimes lies in the clustering
amplitude of the brightest ($M_B < -24$) quasars. Because of the steep   
dependence of the quasar luminosity function on magnitude, there are very few such objects
at low redshifts in 
any flux-limited survey. For example, in the 2dF 10k QSO catalogue (Croom et al 2001a),
which covers an effective area of 290 deg$^{2}$ of the sky,  there
are 815 quasars with $0.3 < z < 0.57$, but only 108 quasars in this
redshift range have B-band magnitudes brighter than  -24. 
However, because the  
mean separation between galaxies and quasars is a factor of ten smaller than the mean
separation between the quasars themselves, 
it will still be possible to obtain a robust measurement of $\xi_{\rm QG}$ for these objects,
particularly if photometric or spectroscopic redshifts are available for the galaxies in
the sample.

\begin{figure}
\centerline{
\epsfxsize=10cm \epsfbox{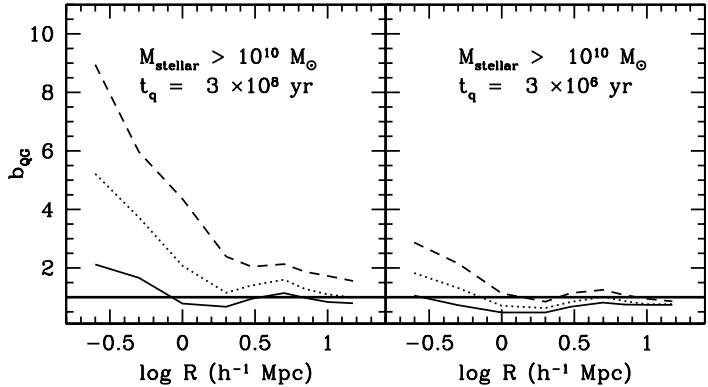}
}
\caption{\label{fig7}
\small
$b_{\rm QG}(R)$ for long and short quasar lifetimes.                               
Solid, dotted and dashed lines show results for quasars with $M_B =$ -22.5,
-23.5 and -24.5. Results are shown at $z=0.4$. }
\end {figure}
\normalsize

\section {Quasar/Galaxy Correlations at Higher Redshift}

Fig. 1 shows that quasars of given optical luminosity are located in lower mass halos at
higher redshifts in the KH model. There are two reasons for this:
1) galaxies are more gas-rich at high redshifts, so that $M_{acc}$ in equation (1) is larger
for a galaxy of given mass at high $z$. 2) the quasar lifetime $t_q$ is shorter at higher
redshift.
Both these assumptions were necessary in order to explain the strong observed
increase in the space densities of bright quasars from the present day out to $z \sim 2$ 
(see KH for a more detailed discussion).

One consequence of these assumptions is that quasars of given
optical luminosity are located in less massive host galaxies at high redshifts.
These predictions appear to have been confirmed in two recent studies (Rix et al 1999;
Ridgway et al 2001), although the results of  Kukula et al (2001) indicate that
quasar hosts at $z \sim 2 $ may be somewhat more massive than predicted by the KH model.

The evolution of $b_{\rm QG}(R)$ to high redshift is shown in Fig.8. 
Our prediction is that the relative bias between quasars
and galaxies {\em decreases} at high redshift. On scales larger than $1 h^{-1}$ Mpc, the predicted
evolution of $b_{\rm QG}$ depends only weakly on galaxy selection.                 
On scales less than $1 h^{-1}$ Mpc, the predicted
evolution of $b_{\rm QG}$ is very sensitive to how galaxies are selected.
The bias decreases more strongly with redshift if galaxies are
selected by stellar mass than if they are selected by                   
rest-frame B-band  magnitude. (Note that a galaxy of fixed B-band magnitude 
has a smaller stellar mass at high redshift, because optical mass-to-light ratios 
increase strongly with age.) 

\begin{figure}
\centerline{
\epsfxsize=10cm \epsfbox{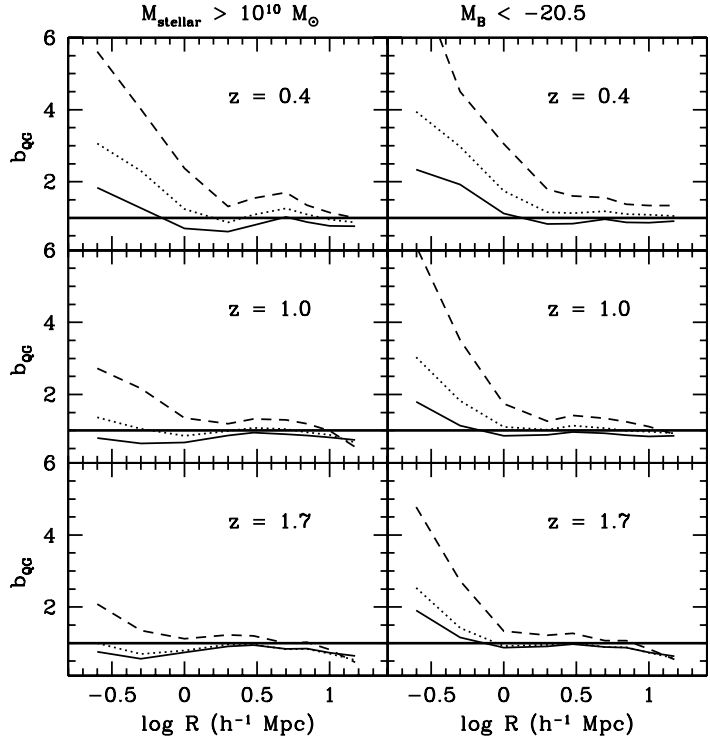}
}
\caption{\label{fig8}
\small
$b_{\rm QG}(R)$ at z=0.4, 1 and 1.7 for galaxies selected according to stellar mass
and according to rest-frame B-band absolute magnitude.
Solid, dotted and dashed lines show results for quasars with $M_B = -22.5$,
$-23.5$ and $-24.5$. Results are shown for the fiducial model with $t_q (z=0) = 3\times 10^7$ yr.}
\end {figure}
\normalsize

\section {Evolution of the Quasar Auto-correlation Function}

We now turn to the predicted evolution of the quasar autocorrelation function $\xi_{\rm QQ}$.
On small scales, the number of quasar pairs will depend strongly on the physical
processes responsible for channelling gas to the central black holes. However, on large
scales, $\xi_{\rm QQ}$ simply probes the mass distribution of dark matter halos
in which quasars are located.

Our model predictions for the evolution of the clustering length $R_0$ as a function of
redshift are shown in Fig. 9 for different choices of $t_q$ (Note that quasar
lifetimes are always quoted at z=0). Results are shown
for quasars with $ -24 <  M_B < -23$ (short-dashed lines) and for brighter quasars with 
$-26 < M_B  < -25$ (long-dashed lines).
Because the mean separation between            
quasars is more than 10 times the mean separation between galaxies, the quasar
auto-correlations are always considerably more noisy than the
quasar/galaxy cross-correlations. In Fig. 10, we plot the evolution
of the co-moving correlation length $R_0$ and compare it with
recent measurements of the redshift-space clustering length $s_0$
from the 2dF QSO survey (Croom et al 2001b).
The sample analyzed by Croom et al (2001a) is flux-limited and is dominated by quasars
with $M_B > -23$ at low redshifts and by quasars with $M_B <  -26$ at $z>2$.
At redshifts $\sim 1$, there are roughly equal numbers of galaxies
in the two magnitude ranges $-24 < M_B < -23$ and $-26 < M_B < -25$. 
To facilitate comparison with the published results, we
have also plotted the evolution of $R_0$ in our simulations adopting the same flux limits
as in the 2dF data (solid lines).  
We only show results over the redshift range where the quasars
in our simulation span the range of absolute magnitudes of the quasars in the 
data ($z=0.6-1.8$).

We find that the data are in better agreement with short  quasar
lifetimes. It is reassuring that the same values of the lifetime 
($10^6-10^7$ years) provide a good fit to the
evolution of the quasar auto-correlation function and to the evolution of 
the quasar luminosity function (Fig. 2) at redshifts  
less than 1. Quasar lifetimes of $10^8$ years are strongly disfavoured by the data.

\begin{figure}
\centerline{
\epsfxsize=10cm \epsfbox{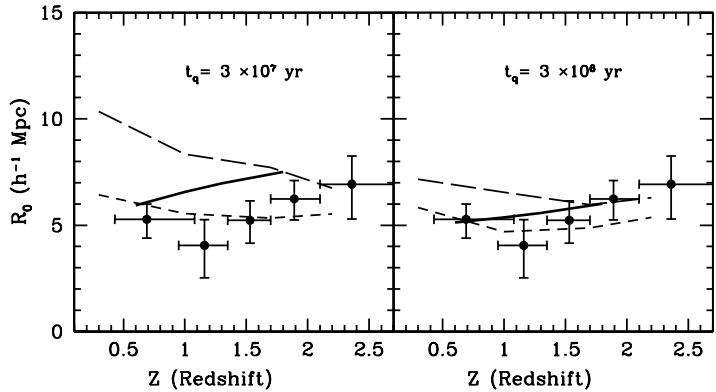}
}
\caption{\label{fig9}
\small
The evolution of the clustering length $R_0$ as a function of redshift for quasars
with $M_B = -23.5$ (short-dashed  lines) and $M_B =-25.5$ ( long-dashed lines) for two
different values of the quasar lifetime at z=0.
The data points are taken from Croom et al (2001b). The solid line shows the
evolution of $R_0$ for quasars with the same distribution of absolute
magnitudes as in the data.}
\end {figure}
\normalsize

\section {Summary and Discussion}

We have studied the quasar/galaxy cross-correlation function $\xi_{\rm QG}$ 
and the quasar/quasar auto-correlation function $\xi_{\rm QQ}$ by embedding models for
 the formation and evolution of the two  populations  in cosmological N-body simulations.
The galaxy evolution model has been described in detail by Kauffmann et al (1999a). 
The quasar evolution model is that of Kauffmann \& Haehnelt (2000) where          
supermassive black holes are formed and fuelled during major mergers.
The correlation functions of quasars are interesting, because they
probe physical processes operating on different scales.

On large scales ($> 1 h^{-1}$ Mpc), 
most of the contribution to $\xi_{\rm QG}$ and $\xi_{QQ}$ comes from
pairs in different dark matter halos.
We define a quasar/galaxy bias parameter $b_{\rm QG}$ as the ratio of the amplitude
of $\xi_{\rm QG}$ to that of $\xi_{\rm GG}$, the galaxy auto-correlation function.
On scales larger than $1 h^{-1}$ Mpc, $b_{\rm QG}$ depends very little on
how galaxies are selected in the simulation. We obtain very similar results if
we select galaxies according to stellar mass or according to 
optical luminosity.

On large scales, $\xi_{\rm QG}$ and $\xi_{QQ}$  both probe 
the mass distribution of the dark matter halos  
that contain quasars. If we adopt the assumptions of the KH model,  
they both provide a measure of the typical quasar lifetime.         
At low redshifts, $\xi_{\rm QG}$    
provides a more efficient way of estimating the quasar lifetime than
$\xi_{\rm QQ}$, because an accurate measurement
of $\xi_{\rm QG}$ is possible for a relatively small number of quasars. 
For the number of quasars available in current surveys, it will be possible to
constrain the dependence of the lifetime on luminosity, on host galaxy-type and on radio power. 
At redshifts greater than $\sim 1$, where it becomes much more difficult to
obtain spectroscopic redshifts for galaxies and where the space density of quasars
is considerably higher, the quasar auto-correlation function $\xi_{\rm QQ}$ 
will probably prove more useful.
We have used the simulations to  calculate the evolution 
of $\xi_{\rm QQ}$.                            
Our models agree with the results of the 2dF QSO survey for quasar lifetimes 
in the range $10^{6}-10^{7}$ years.

On scales less than $1 h^{-1}$ Mpc, 
$\xi_{\rm QG}$ and $\xi_{\rm QQ}$ are dominated
by  pairs within the same dark matter halo. 
The slope and  amplitude of $\xi_{\rm QG}$ on these scales depend
not only on                           
the  number and the distribution of galaxies in the halo, but on the location of
the quasar in the halo. In principle, it should be possible to constrain  
the density profiles of galaxies in dark matter halos directly from the observations.
When this  is combined with the clustering amplitude of quasars 
measured on large scales,
it should be possible to test whether quasars are located at
the  centres of dark matter potential wells, or whether they are more
uniformly distributed throughout the halo. This would yield important information
about the physical processes that were responsible for forming and fuelling the
supermassive black holes found at the centres of galaxies today.

\vspace{0.8 cm}
\large
{\bf Acknowledgments}\\
\normalsize
We thank Brian Boyle and Scott Croom for helpful advice and for providing their
data in machine readable format. This work was supported by the European 
Community Research and Training Network "The Physics of the Intergalactic
Medium". GK thanks the Astrophysics Group of Imperial College for their 
hospitality. 

\pagebreak 
\Large
\begin {center} {\bf References} \\
\end {center}
\normalsize
\parindent -7mm  
\parskip 3mm

Bahcall J.N., Schmidt M., Gunn J.E., 1969, ApJ, 157, L77                        

Bahcall J.N., Chokshi A., 1991, ApJ, 380, L9

Brown M.J.I., Boyle B.J., Webster R.L., 2001, AJ, 122, 26

Croom S.M., Smith R.J., Boyle B.J., Shanks T., Loaring N.S., Miller L.,
Lewis I.J., 2001a, MNRAS, 322, L29

Croom S.M., Shanks T., Boyle B.J., Smith R.J., Miller L., Loaring N.S.,
Hoyle F., 2001b, MNRAS 325, 483

Diaferio A., Kauffmann G., Balogh M.L., White S.D.M., Schade D., Ellingson E.,
2001, MNRAS, 323, 999

Ellingson E., Green R.F., Yee H.K.C., 1991, ApJ, 371, 49

Finn R.A., Impey C.D., Hooper E.J., 2001, ApJ, in press (astro-ph/0104124)

Fisher K.B., Bahcall J.N., Kirhakos S., Schneider D.P., 1996, ApJ, 468, 469

Haehnelt M.G., Rees M.J., 1993, MNRAS, 263, 168

Haehnelt M.G., Natarajan P.,  Rees M.J., 1998, MNRAS, 300, 817

Haehnelt M.G., Kauffmann G., 2000, MNRAS, 318, L35

Haiman Z.,  Hui L., 2001, ApJ, 547, 27

Jing Y., 1998, ApJ, 503, 9

Kauffmann G., White S.D.M., 1993, MNRAS, 261, 921

Kauffmann G., Nusser A., Steinmetz M., 1997, MNRAS, 323, 999

Kauffmann G., Colberg J., Diaferio A., White S.D.M., 1999a, MNRAS,303,188 

Kauffmann G., Colberg J., Diaferio A., White S.D.M., 1999b, MNRAS,307,529 

Kauffmann G., Haehnelt M., 2000, MNRAS, 311, 576

Kochanek C.S., Falco E.E., Munoz J.A., 1999, ApJ, 510, 590

Kukula M.J., Dunlop J.S., McLure R.J., Miller L., Percival W.J., Baum S.A.,
O'Dea C.P., 2001, MNRAS, in press (astro-ph/0010007)

La Franca F., Andreani P., Cristiani S., 1998, ApJ, 497, 529

Martini P., Weinberg D.H., 2001, ApJ, 547 12

McLure R.J., Kukula M.J., Dunlop J.S., Baum S.A., O'Dea C.P., Hughes D.H., 1999,
MNRAS 308, 377

McLure R.J., Dunlop J.S., 2001, MNRAS, 321, 515

Mo H.J., White S.D.M., 1996, MNRAS, 282, 347

Ridgway S.E., Heckman T.M., Calzetti D., Lehnert M., 2001, ApJ, 550, 122

Rix H.-W., Falco E. Impey C., Kochanek C., Lehar J., McLeod B.A., Munoz J.,
Peng C., 1999 (astro-ph/9910190)

Sheth R.K., Mo H.J., Tormen G., 2001, MNRAS, 323, 1

Springel V., White S.D.M., Tormen G., Kauffmann G., 2001, MNRAS, 
 in press (astro-ph/0012055)

Wold M., Lacy M., Lilje P.B., Serjeant S., 2001, MNRAS, 323,231

York D.G., Adelman J., Anderson J.E., Anderson S.F., Annis J., Bahcall N.A., Bakken J.A.,
Barkhouser R. et al., 2000, AJ, 120, 1579

Yee H.K.C., Green R.F., 1984, ApJ, 280, 79

Yee H.K.C., Green R.F., 1987, ApJ, 319, 28

\end {document}